\documentclass[conference]{IEEEtran}
\IEEEoverridecommandlockouts
\usepackage{cite}
\usepackage{amssymb,amsfonts}
\usepackage{algorithmic}
\usepackage{textcomp}
\usepackage{xcolor}
\usepackage{amsmath,graphicx}
\usepackage{booktabs}
\def\BibTeX{{\rm B\kern-.05em{\sc i\kern-.025em b}\kern-.08em
    T\kern-.1667em\lower.7ex\hbox{E}\kern-.125emX}}
\begin{document}

\title{Adapitch: Adaption Multi-Speaker Text-to-Speech Conditioned on Pitch Disentangling with Untranscribed Data
}


\author{\IEEEauthorblockN{Xulong Zhang, Jianzong Wang$^\ast$\thanks{$^\ast$Corresponding author: Jianzong Wang, jzwang@188.com.}, Ning Cheng, Jing Xiao}
\IEEEauthorblockA{\textit{Ping An Technology (Shenzhen) Co., Ltd.}}
}

\maketitle

\begin{abstract}
    In this paper, we proposed Adapitch, a multi-speaker TTS method that makes adaptation of the supervised module with untranscribed data. We design two self supervised modules to train the text encoder and mel decoder separately with untranscribed data to enhance the representation of text and mel. To better handle the prosody information in a synthesized voice, a supervised TTS module is designed conditioned on content disentangling of pitch, text, and speaker. The training phase was separated into two parts, pretrained and fixed the text encoder and mel decoder with unsupervised mode, then the supervised mode on the disentanglement of TTS. Experiment results show that the Adaptich achieved much better quality than baseline methods. 
\end{abstract}


\section{Introduction}
\label{sec:intro}

 Early speech synthesis methods mainly include concatenative synthesis and parametric synthesis~\cite{abs-2106-15561}. The concatenative synthesis builds a huge database for different phoneme speech audio, for the specific sentence it retrieval and concat the phoneme voice sequence together for the output. While the building of a speech database is complicated, the synthesized voice is natural of the human. The parametric synthesis is to relieve the directly concated from waveform, it generates acoustic parameters to reconstruct the waveform. The parametric synthesis has low data cost, but it is heard as robotic and can be easily recognized as not human speech.

Recently, neural network based text-to-speech has achieved great success, allowing us to synthesize more natural and realistic speech~\cite{Zhao_nnspeech,ren_fastspeech_2019}. In addition to the demand for the naturalness of speech synthesis, more and more personalized speech synthesis needs require higher speech quality at the same time. However, in many practical application scenarios, only a small amount of speech data of the target speaker can be obtained, and personalized speech synthesis for the target speaker is still a big challenge. This task is the so-called low-resource TTS that only requires a small amount of reference corpus for speech synthesis.

In order to solve the low-resource TTS task, previous studies~\cite{liu_towards_2020,zhang2022TDASS} have proposed a method of transfer learning adaptation. First, train a complete TTS model through speaker data with a large amount of corpus, and then use the corpus of the target speaker to fine-tune the trained model. The problem with this method is that a certain amount of reference corpus is required for fine-tuning. The model after fine-tuning has a certain degree of loss in naturalness and speech quality in speech synthesis. In addition, the fine-tune method cannot be applied to the scene of real time speech synthesis, making this method unattractive.

In addition to the fine-tune method, previous studies~\cite{huybrechts2021low,zheng_improving_2020-1,tang2022avqvc} have proposed the use of speaker embedding. During training, speaker embedding is added for joint training to avoid using target data to fine-tune the model. However, accents are often mismatched and nuances such as characteristic prosody are lost. It is difficult to find a single feature to represent all the voice data and pronunciation style of the target speaker.

In order to solve the problem of speaker embedding features, some studies~\cite{eskimez_end--end_2020,wagner2010experimental,zhang2022SUSing} have proposed embedding a variety of features other than speaker features. Each feature represents different attributes of the speaker corpus, such as prosody embedding and style embedding. But this method only combines more single dimensions to characterize the speech synthesis of the target speaker. A small number of features obtained from fewer target data means that it is difficult to cover all the pronunciation of the target speaker.

We utilize untranscribed data to pretrain the pluggable text encoder and mel decoder in self supervised mode, with a supervised module for disentangling the content of speech to obtain pitch and speaker embedding. It includes the encoding of phoneme sequence, the variance adaptor module to disentangle pitch, text, and speaker, and the decoding module to get the final mel spectrum. Pitch regressor and speaker Look Up Table (LUT) are introduced in the disentangle module. The LUT is used to control the encoding to enhance the different speaker feature embedding. The pitch regressor is used to control the pitch information in the speech synthesis that is independent. The encoder of the phoneme controls the content without the different speaker related features. The mel decoder through a self supervised training to keep robust decoding of mel spectrum.

Our contributions are as follows:
1) Two pluggable self supervised submodules are designed to utilize the untranscribed data for enhancing the representation of text encoder and mel decoder.
2) An adaption between supervised input and self supervised input of mel decoder are added to utilizing the pretrained mel decoder.

\section{Method}

\subsection{Model Overview}
\label{method}

\begin{figure}[htp]
	\includegraphics[width=1.0\linewidth]{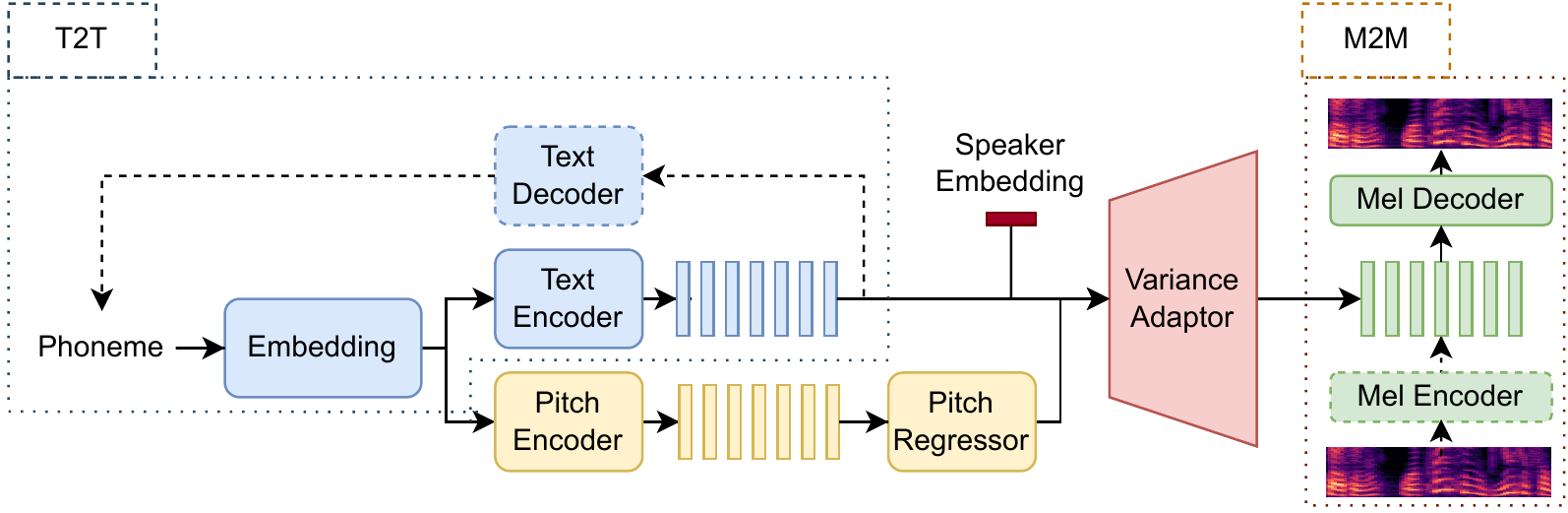}
	\caption{The overview of the proposed text-to-speech method.}
	\label{fig:1  overview}
\end{figure}
As shown in Figure~\ref{fig:1  overview}, it is the architecture of our method based on Fastspeech2~\cite{ren_fastspeech_2021}. The difference is the introduced self supervised of text to text module (T2T) and mel to mel module (M2M). To be noted, in the T2T module the text is in the representation of phoneme. There are two separated phases for training the proposed model. One is the self supervised training, it contains phoneme to phoneme and mel spectrogram to mel spectrogram, namely the M2M module. From the training of phoneme to phoneme, namely T2T module, to get a speaker independent content representation of the input text. And with utilizing the mel spectrogram to mel spectrogram training could use untranscribed data to enhance the decoder phase. The second is the supervised training on the whole model with the pretrained parameters of text encoder and mel decoder. There is an adaptation between the supervised mode input of mel decoder and self supervised mode input of mel decoder.

\subsection{Text to Text Module (T2T)}
As depicted in the left of Figure \ref{fig:1  overview}, the T2T consists of an embedding layer, text encoder, and text decoder. The module of T2T mainly conducts the representation of the text from the format of the phoneme. The learned latent variable should reconstruct the content of text but without the information of speaker timbre information from the mel spectrum along with the backpropagation of gradient. To avoid the affect of the effect from the target mel spectrum, an aux task of text reconstruction is added by introducing a text decoder. A text decoder was added to the T2T module to do a reconstruction of the phoneme. The text reconstruction task also can resolve the lack of supervised data of the target speaker, the ability of the text encoder could be enhanced by data augmentation. In the T2T module, we utilize many pure text sentences for the self-supervised training to enable the text encoder to learn the representation of all the phonemes robustly. The text encoder was built up with three conv1d layers and a bi-LSTM layer. The text decoder includes an LSTM layer, an attention layer, and the linear norm layer. The text embedding with the size of 512, the kernel size of conv1d is 5, stride is 1, the activation function is relu, and the hidden layer size of bi-LSTM is 256. 

\subsection{Mel to Mel Module (M2M)}
As shown in the right of Figure~\ref{fig:1  overview}, the M2M module is to train a mel decoder which enables decode the latent variable from mel encoder and the full connection layer into mel spectrum. The M2M module is an encoder-decoder architecture, it consists of a mel encoder and mel encoder. There are two phases of training for the M2M module. One is the self-supervised training to do the reconstruction task on the untranscribed speech. The other one is the supervised training, with fixed parameters of the mel encoder and mel decoder, the output of the mel encoder was used as a label for the full connect layer, which forces the adjustment of the latent variable representation of the encoder. The training of M2M makes the supervised representation of the acoustic feature can be guided by the mel encoder. The mel encoder could be a teacher for the supervised data to learn the feature representation of the mel spectrum. The mel encoder is built up by a three-layer stacked convolutional neural network, in which the kernel size is (3,3). The mel decoder consists of 4 layers of feed forward transformer.

\subsection{Variance Adaptor}
The text encoder and mel decoder networks are designed based on the transformer module, and it has achieved effective results in speech synthesis tasks. In addition to the T2T module to process the phoneme and M2M module to obtain the mel spectrum, our proposed model adds a variance adaptor module between the text encoding and mel decoding. The process of variance adaptor mainly involves the decoupling of pitch and text content and the enhancement of speaker identity. Corresponding pitch regression and look up table (LUT) for distinguishing multiple speakers are designed. In the duration expansion of speaker embedding, latent variable of the output of text encoder and pitch values, linear interpolation and nearest neighbor interpolation are used. Through upsampling, they have the same size to stack.

\subsection{Training and Inference}

In the model training process, the phoneme sequence was fed to the embedding layer, and then the transformer layers build up the text encoder and the pitch encoder to obtain the latent variables. For the obtained latent variables, the corresponding pitch sequence value will be generated by pitch regression and compared with the real pitch value. 
The process of pitch regression can help the latent variable of the encoder to contain features that can represent pitch information. By optimizing pitch regression, the encoder's ability to express pitch features can be improved. The LUT records the representations of different speakers and is used to distinguish phonemes and voice pairs of different speakers during the training process. In the training process, the actual pitch and outputs of the encoder and the speaker embedding information from the LUT are concated according to the time extension. The fused embedding is fed to the fully connected layer and then transformer based mel decoder to obtain the mel spectrum corresponding to the corpus. The output of the variance adaptor is tuned close to the latent variables of the mel encoder. 

In the model inference process, the encoder module drops the text decoder and keeps the text encoder for content embedding and duration predictor. The pitch encoder produces the latent variables for pitch regression. The decoder part only keeps the mel decoder. In the variance adaptor, the main difference is the predicted pitch instead of the real pitch in the training phase. 


\subsection{Training Loss}

To describe our model more formally, let $E_T(\cdot)$ and $E_P(\cdot)$ stand for the text encoder and pitch encoder separately. Let $D_T(\cdot)$ and $E_M(\cdot)$ be the text decoder and Mel encoder, $D_p(\cdot)$ denote the duration predictor, $R_p(\cdot)$ be the pitch regressor. Let $t_i$ be the $i$th phoneme of the input, $v_j$ be the embedding of the speaker $j$, $s_j$ be the target audio of speaker $j$. And $j$ denote a speaker. The output of the model can be written as Equation (\ref{eq1:out model}):
\begin{equation}
\label{eq1:out model}
F(t_i,j)=D_M(FC(\left[
\begin{array}{c}
E_T(t_i)\\
R_p(s_j)\\
v_j\\
\end{array}
\right]))
\end{equation}
The $D_M(\cdot)$ is the Mel spectrum decoder, $FC(\cdot)$ is the fully connected layer. There is a synthesis loss as Equation~(\ref{eq2: synthesis loss}):
\begin{equation}
\label{eq2: synthesis loss}
\mathcal{L}_{syn}=\sum_i\sum_{s_j}\mathcal{L}_{mse}(F(t_i,j),mel(s_j))
\end{equation}
where the $mel(s_j)$ is the Mel spectrum of the target audio $s_j$. For the pitch regression module, there is a regression loss as shown in Equation (\ref{eq3: reg loss}):
\begin{equation}
\label{eq3: reg loss}
\mathcal{L}_{reg}=\sum_i\sum_{s_j}\mathcal{L}_{mse}(R_p(E_P(t_i)),P(s_j))
\end{equation}
where $P(\cdot)$ is the function of pYin~\cite{mauch2014pyin} to estimate the pitch of the target audio. 

There are two self-supervised pretrained modules in the proposed model. The text-to-text module trains the text encoder to keep speaker-independent. There is a reconstruct loss of the text $\mathcal{L}_{rec}^T$, as shown in Equation~(\ref{eq:text recon loss}):

\begin{equation}
\label{eq:text recon loss}
\mathcal{L}_{rec}^T=\sum_i\mathcal{L}_{mse}(D_T(E_T(t_i)),t_i)
\end{equation}
where $D_T(\cdot)$ is the text decoder. The other self-supervised module is the mel to mel, it can enlarge the dataset with untranscribed speech. There is also a reconstruct loss $\mathcal{L}_{rec}^M$ for the mel spectrum, as show in Equation~(\ref{eq:mel recon loss}):

\begin{equation}
\label{eq:mel recon loss}
\mathcal{L}_{rec}^M=\sum_k\mathcal{L}_{mse}(D_M(E_M(m_k)),m_k)
\end{equation}
where $m_k$ is the $k$th frame mel spectrum. 

Due to the proposed method try to utilize the untranscribed data to enhance the Mel decoder, the latent variables out from the variance adaptor should be consistent with the output of Mel encoder. There is an adaptation loss $\mathcal{L}_{ada}$ to pull the two latent variables close, as depicts in Equation~(\ref{eq:ada loss}):

\begin{equation}
\label{eq:ada loss}
\mathcal{L}_{ada}=\mathcal{L}_{mse}(E_M(\mathbb{M}),FC(\left[
\begin{array}{c}
E_T(\mathbb{T})\\
R_p(\mathbb{S})\\
v_j\\
\end{array}
\right]))
\end{equation}
where the $\mathbb{T}$ represents the input phonemes, the $\mathbb{S}$ represent the speech, and $\mathbb{M}$ represents the Mel spectrum.

The total loss is the combination of the synthesis loss $\mathcal{L}_{syn}$, regression loss $\mathcal{L}_{reg}$ and adaptation loss $\mathcal{L}_{ada}$ as Equation (\ref{eq 5: total loss}):
\begin{equation}
\label{eq 5: total loss}
\mathcal{L}_{total}=\alpha\mathcal{L}_{syn}+\beta\mathcal{L}_{reg}+\gamma\mathcal{L}_{ada}
\end{equation}
where $\alpha$,$\beta$,$\gamma$ are three hyperparameters. During the training phase, we minimize $\mathcal{L}_{ada}$ and $\mathcal{L}_{syn}$ to optimize mel spectrum generation. Minimize $\mathcal{L}_{reg}$ for pitch regressor. Finally, the whole model was optimized by using $\mathcal{L}_{total}$.

\section{Experiments}
\label{experiments}
In the comparison, we mainly choose the Fastpitch as the baseline method for pitch estimated, which is implemented from the opensource code and rerun with the same dataset. We also compared the performance of synthesized voice with the popular TTS methods of Tacotron2~\cite{tacotron2}, Fastspeech2~\cite{ren_fastspeech_2019} and Transformer TTS~\cite{li2019neural}.



\subsection{Datasets}
In the experiment, we used three datasets and separately pretrain the two sub-modules with different datasets. We selected a public dataset of LibriTTS~\cite{zen2019libritts} for the M2M module pretraining, only the audio was used and the transcribed text was discarded. The LibriTTS has a total of 585 hours of speech, and it contains 2456 speakers. To pretrain the T2T module, we choose another public dataset of VCTK~\cite{vctk}. The VCTK dataset contains 108 speakers were used, each speaker read about 400 sentences. The total size of the VCTK is about 44 hours. During the pretraining of the T2T module, only the transcribed text data of the audio in VCTK were used, and the audio was discarded. The final proposed model is trained on a small public dataset of LJSpeech~\cite{ljspeech17} in a supervised manner, both audio and text are used. 

\subsection{Training and inference}
For the model training, we need to use a text corpus for the enhancement of the text understanding in the text encoder. For the mel decoding from the latent vector, we set a self reconstruct task to do a mel encoder and mel decoder process. The mel decoder was trained on a speech corpus. With the pretrained text encoder and mel decoder, we conduct supervised training of the module of pitch regressor and speaker variance adaptor. The generated latent vector is also supervised by the output of the mel encoder. For the target speaker, we do the adaptation on the supervised module finetune and keep the module of mel encoder, mel decoder, and text encoder. To be noted the ground truth pitch value instead of the output of pitch regressor was used for the input of variance adaptor.To be summarized, there are two stages during the training phase:
1) self-supervised training for the text to text module and the mel to mel module separately.
2) with the pretrained sub-modules to train the supervised model for text to mel condition on the disentangling of text embedding, speaker embedding and pitch embedding.

Thus the pretrained sub modules could be pluggable for neural TTS with other network architecture to enhance the text encoder and mel decoder. 

The inference was conducted without the text decoder module and the mel encoder module. Finally, with the generated mel spectrum through a pretrained WaveGlow model for the wave audio synthesis.

\subsection{Analysis of the Pitch of Synthesized Audio}
The pitch is the main component of the speech, we set a pitch encoder to enhance the learning of pitch information. To validate it do helps the synthesis speech, we calculated the pitch extracted from the synthesized speech by our model and the comparison speeches from Fastpitch and groundtruth. In order to evaluate the extract pitch for the comparision of synthesised speech and the groundtruth, two metrics about pitch extraction is introduced from the work of~\cite{BabacanDDHD13}. They are Gross Pitch Error (GPE) and Fine Pitch Error (FPE). The GPE is the relative pitch error with a threshold, compaired with the groundthruth for the voiced part. The FPE is the standard deviation of the distribution  of relative error values (in cent) from the frames without gross pitch errors. The mean square error of pitch value of the synthesized speech between the groundtruth was also calculated on the test dataset of LJSpeech, the results are shown in Table~\ref{table:pitch mse}.  

\begin{table}[h]
    \centering
    \caption{The evaluation metrics of pitch between synthesis speeches and groundtruth on the test set of LJSpeech}
    \begin{tabular}{cccc}
        \toprule
         Method& Mean Square Error (\%)&FPE (cents)&GPE (\%)\\
         \midrule
     Fastpitch~\cite{fastpitch2021}&35$\pm$4 &14.63&2.46\\
     Adapitch&23$\pm$6 &17.69&1.28\\
         \bottomrule
    \end{tabular}
    
    \label{table:pitch mse}
\end{table}

\begin{figure}[htp]
\centering
	\includegraphics[width=0.8\linewidth]{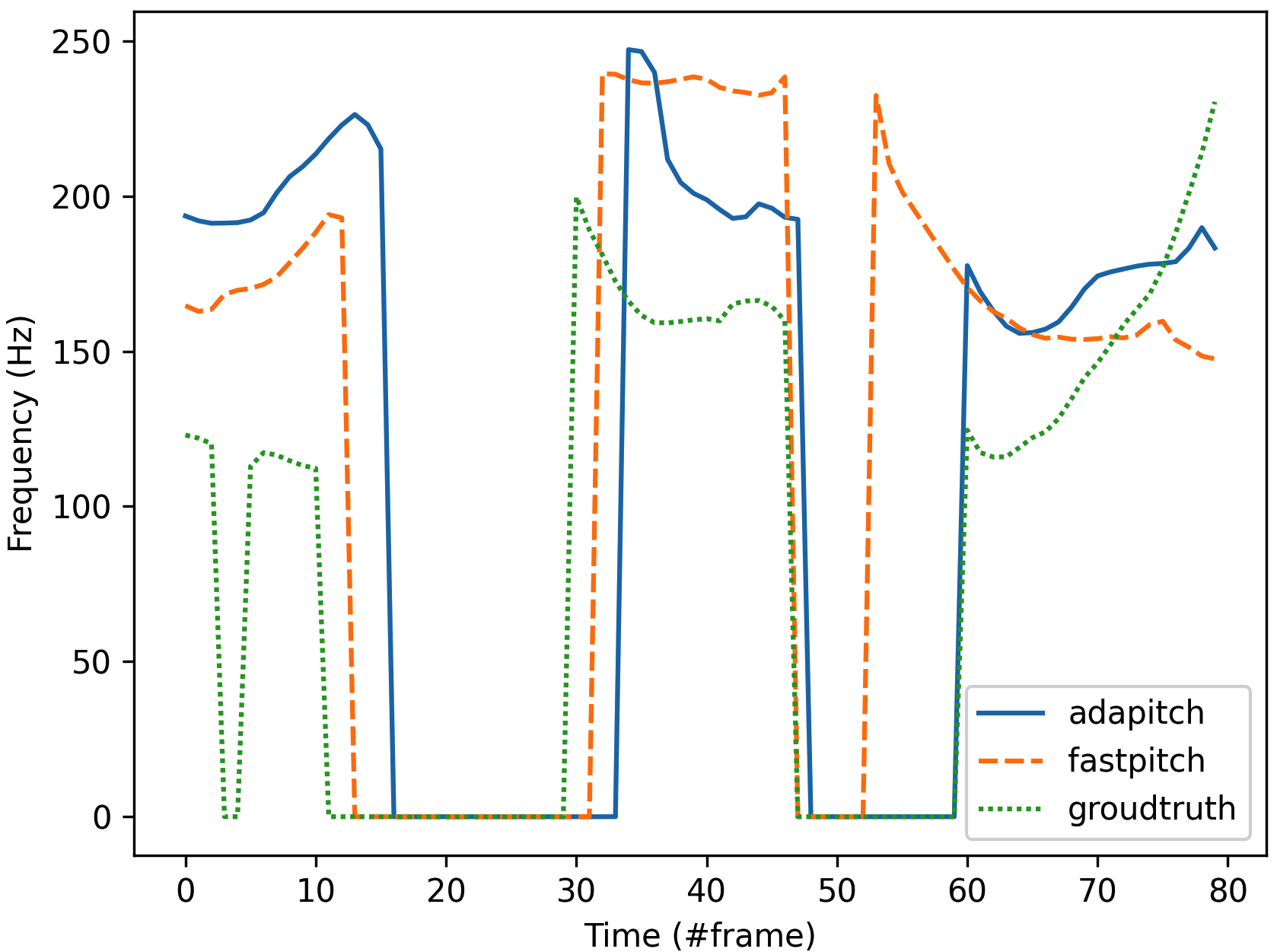}
	\caption{Comparison of synthesized audio pitch visualization, the pitch estimated by pYin and the audio clip is 80 frames.}
	\label{fig:2 pitch}
\end{figure}

The method of Fastpitch had the same pitch representation as our method, but without the enhancement of the text encoder. As the results are shown in Table~\ref{table:pitch mse}, the proposed method of Adapitch has lower MSE than the method of Fastpitch, Adapitch achieves lower gross error rates than the baseline method of Fastpitch, and also achieve a higher FPE value. We can conclude that the unsupervised T2T module helps the pitch embedding for the pitch representation in the text. We also visualized the pitch of a clip of synthesized audio with groundtruth in Figure~\ref{fig:2 pitch}. Our proposed method Adapitch labeled with a blue solid line and the baseline method Fastpitch in comparison is labeled with an orange dotted line. 

As depicted in Figure~\ref{fig:2  pitch}, the pitch of Fastpitch is smoothly in the time between frame 30 and frame 50, whereas the pitch of Adapitch could keep variety as the pitch of the ground truth. Especially during the frame of 60 to 80, the pitch of Fastpitch gives an opposite trend with the ground truth. From the pitch comparison, we can conclude that the pitch values of the proposed method are closer to the ground truth. 

\subsection{Subject Listening Test }

We conducted a listening test with a total of 5 sentences for each model to synthesize the audio. We included nine systems in the listening test: groundtruth (audio), groundtruth mel + Vocoder (with raw spectrum synthesis audio by WaveGlow), Tacotron2, Fastspeech2, Transformer TTS and Fastpitch as comparison and the three proposed systems. In the listening test, we have 12 persons to give a 5-point scale for each audio, the value of 1 means the quality of the audio is very poor, and 5 means the quality of the audio is perfect. The mean opinion scores with 95\% confidence intervals are listed in Table~\ref{table compare mos}. To show that the self supervised sub modules (T2T and M2M) have effects on the enhancement of the text encoder and mel decoder, we conducted the ablation experiment and train the different models separately.

\begin{table}[htbp]   \caption{\label{table compare mos}Mean Opinion Scores (MOS) with 95\% confidence intervals and Mel Cepstral Distortions (MCD). }  
	\centering
	\begin{tabular}{lcl}   
		\toprule 
		Method & MOS & MCD\\
		\midrule 
			Groundtruth & 4.80$\pm$0.03&-\\
		Groundtruth mel + Vocoder & 4.73$\pm$0.04&-\\
		Tacotron2~\cite{tacotron2}& 3.71$\pm$0.03&8.12\\
		Fastspeech2~\cite{ren_fastspeech_2019}& 3.66$\pm$0.03&8.23\\
		Transformer TTS~\cite{li2019neural}& 3.77$\pm$0.03&7.53\\
		Fastpitch~\cite{fastpitch2021} & 3.80$\pm$0.04&8.28\\
		\midrule
		Adapitch w/o M2M & 3.73$\pm$0.03&7.83\\
		Adapitch w/o T2T & 3.60$\pm$0.04&9.25\\
		Adapitch & 3.93$\pm$0.04&7.72\\
		\bottomrule   \end{tabular}  
\end{table}

As shown in the results of MOS test, the score of groundtruth and the ground truth mel + vocoder achieved nearly 5 points. It shows that WaveGlow has slightly decreased the quality of audio. When comparing the results of the ablation experiment, without the T2T module decreased 0.33 and without the M2M module decreased 0.20. Both the text encoder and the mel decoder have an impact on the proposed method. Finally, compared with the baseline methods, it outperforms the popular method of Tacotron2, Fastspeech2, and Transformer TTS. It shows 0.13 improvement was achieved over the best of Fastpitch.

\subsection{Objective Evaluation}

To compare the synthesized audio objectively, we conduct an evaluation using mel cepstral distortions (MCD). We also test several hyperparameters in the experiment to make a selection on the loss weight of the $\alpha$,$\beta$,$\gamma$. As the training of the model is time consuming, grid search for all the hyperparameters is not impossible. We try several parameter combinations and use the MCD between the ground truth to decide the adjusted trend of the hyperparameters. We keep the $\alpha$ as 1 and tune the $\beta$ and $\gamma$. With the $\beta$ increase from 0.1 to 2, the MCD value varied slightly. On the tune of $\gamma$ also have little change in terms of MCD value. This result shows that the audio synthesized by different loss weights is not much different after the model training converges. During the training phase, with the increase of $\gamma$ the need epochs increase too. But with the increase of $\beta$, the need epochs for training nearly do not change. Finally, we fixed the hyperparameters of $(\alpha,\beta,\gamma)$ as $(1,0.1,0.1)$ for our proposed method of Adapitch.


As shown in Table~\ref{table compare mos}, from the comparison with the baseline methods, the Adapitch achieved 7.72 in terms of MCD. It is just higher than the Transformer TTS by 0.19. The lower value MCD shows that the audio is closer to the ground truth. We also find that the score of MCD is slightly different from the MOS value, it may be caused by the quality of synthesized audio too similar to give a precision MOS by subject.


\section{Conclusion}
\label{conclude}

In this work, we proposed Adapitch utilized untranscribed data for self supervised train of the T2T and M2M modules. The text encoder of T2T module and the mel decoder of M2M module were pluggable applied as pretraind models for the supervised text to the mel module. Additionally, in the text to mel module, we conduct a disentanglement on the pitch, text content, and speaker. An adaption between supervised input and self supervised input of mel decoder is designed. Experimental results of the MOS test show the proposed Adapitch has improved than the baseline methods. 

\section{Acknowledgement}
This paper is supported by the Key Research and Development Program of Guangdong Province under grant No.2021B0101400003. Corresponding author is Jianzong Wang from Ping An Technology (Shenzhen) Co., Ltd (jzwang@188.com).

\bibliographystyle{IEEEtran}
\bibliography{mybib}

\begin{thebibliography}{10}
\providecommand{\url}[1]{#1}
\csname url@samestyle\endcsname
\providecommand{\newblock}{\relax}
\providecommand{\bibinfo}[2]{#2}
\providecommand{\BIBentrySTDinterwordspacing}{\spaceskip=0pt\relax}
\providecommand{\BIBentryALTinterwordstretchfactor}{4}
\providecommand{\BIBentryALTinterwordspacing}{\spaceskip=\fontdimen2\font plus
\BIBentryALTinterwordstretchfactor\fontdimen3\font minus
  \fontdimen4\font\relax}
\providecommand{\BIBforeignlanguage}[2]{{%
\expandafter\ifx\csname l@#1\endcsname\relax
\typeout{** WARNING: IEEEtran.bst: No hyphenation pattern has been}%
\typeout{** loaded for the language `#1'. Using the pattern for}%
\typeout{** the default language instead.}%
\else
\language=\csname l@#1\endcsname
\fi
#2}}
\providecommand{\BIBdecl}{\relax}
\BIBdecl

\bibitem{abs-2106-15561}
X.~Tan, T.~Qin, , and et~al., ``A survey on neural speech synthesis,''
  \emph{CoRR}, vol. abs/2106.15561, 2021.

\bibitem{Zhao_nnspeech}
B.~Zhao, X.~Zhang, J.~Wang, N.~Cheng, and J.~Xiao, ``nnspeech: Speaker-guided
  conditional variational autoencoder for zero-shot multi-speaker
  text-to-speech,'' in \emph{{IEEE} International Conference on Acoustics,
  Speech and Signal Processing}, 2022.

\bibitem{ren_fastspeech_2019}
Y.~Ren, Y.~Ruan, and et~al., ``Fastspeech: fast, robust and controllable text
  to speech,'' in \emph{Proceedings of the 33rd International Conference on
  Neural Information Processing Systems}, 2019, pp. 3171--3180.

\bibitem{liu_towards_2020}
A.~H. Liu, T.~Tu, H.-y. Lee, and L.-s. Lee, ``Towards {Unsupervised} {Speech}
  {Recognition} and {Synthesis} with {Quantized} {Speech} {Representation}
  {Learning},'' in \emph{2020 {IEEE} {International} {Conference} on
  {Acoustics}, {Speech} and {Signal} {Processing}}, 2020, pp. 7259--7263.

\bibitem{zhang2022TDASS}
X.~Zhang, J.~Wang, N.~Cheng, and J.~Xiao, ``Tdass: Target domain adaptation
  speech synthesis framework for multi-speaker low-resource tts,'' in
  \emph{2022 International Joint Conference on Neural Networks (IJCNN)}.\hskip
  1em plus 0.5em minus 0.4em\relax {IEEE}, 2022, pp. 1--7.

\bibitem{huybrechts2021low}
G.~Huybrechts, T.~Merritt, and et~al., ``Low-resource expressive text-to-speech
  using data augmentation,'' in \emph{2021 IEEE International Conference on
  Acoustics, Speech and Signal Processing}, 2021, pp. 6593--6597.

\bibitem{zheng_improving_2020-1}
Y.~Zheng, X.~Li, F.~Xie, and L.~Lu, ``Improving {End}-to-{End} {Speech}
  {Synthesis} with {Local} {Recurrent} {Neural} {Network} {Enhanced}
  {Transformer},'' in \emph{2020 {IEEE} {International} {Conference} on
  {Acoustics}, {Speech} and {Signal} {Processing}}, 2020, pp. 6734--6738.

\bibitem{tang2022avqvc}
H.~Tang, X.~Zhang, J.~Wang, N.~Cheng, and J.~Xiao, ``Avqvc: One-shot voice
  conversion by vector quantization with applying contrastive learning,'' in
  \emph{2022 IEEE International Conference on Acoustics, Speech and Signal
  Processing (ICASSP2022)}.\hskip 1em plus 0.5em minus 0.4em\relax IEEE, 2022,
  pp. 4613--4617.

\bibitem{eskimez_end--end_2020}
S.~E. Eskimez, R.~K. Maddox, C.~Xu, and Z.~Duan, ``End-{To}-{End} {Generation}
  of {Talking} {Faces} from {Noisy} {Speech},'' in \emph{2020 {IEEE}
  {International} {Conference} on {Acoustics}, {Speech} and {Signal}
  {Processing}}, 2020, pp. 1948--1952.

\bibitem{wagner2010experimental}
M.~Wagner and D.~G. Watson, ``Experimental and theoretical advances in prosody:
  A review,'' \emph{Language and cognitive processes}, vol.~25, no. 7-9, pp.
  905--945, 2010.

\bibitem{zhang2022SUSing}
X.~Zhang, J.~Wang, N.~Cheng, and J.~Xiao, ``Susing: Su-net for singing voice
  synthesis,'' in \emph{2022 International Joint Conference on Neural Networks
  (IJCNN)}.\hskip 1em plus 0.5em minus 0.4em\relax {IEEE}, 2022, pp. 1--7.

\bibitem{ren_fastspeech_2021}
Y.~Ren, C.~Hu, X.~Tan, T.~Qin, S.~Zhao, Z.~Zhao, and T.~Liu, ``Fastspeech 2:
  Fast and high-quality end-to-end text to speech,'' in \emph{9th International
  Conference on Learning Representations}, 2021.

\bibitem{mauch2014pyin}
M.~Mauch and S.~Dixon, ``pyin: A fundamental frequency estimator using
  probabilistic threshold distributions,'' in \emph{2014 IEEE International
  Conference on Acoustics, Speech and Signal Processing}, 2014, pp. 659--663.

\bibitem{tacotron2}
J.~Shen, R.~Pang, and et~al., ``Natural tts synthesis by conditioning wavenet
  on mel spectrogram predictions,'' in \emph{2018 IEEE International Conference
  on Acoustics, Speech and Signal Processing}, 2018, pp. 4779--4783.

\bibitem{li2019neural}
N.~Li, S.~Liu, and et~al., ``Neural speech synthesis with transformer
  network,'' in \emph{Proceedings of the AAAI Conference on Artificial
  Intelligence}, vol.~33, 2019, pp. 6706--6713.

\bibitem{zen2019libritts}
H.~Zen, V.~Dang, R.~Clark, Y.~Zhang, R.~J. Weiss, Y.~Jia, Z.~Chen, and Y.~Wu,
  ``Libritts: {A} corpus derived from librispeech for text-to-speech,'' in
  \emph{20th Annual Conference of the International Speech Communication
  Association}, 2019, pp. 1526--1530.

\bibitem{vctk}
J.~Yamagishi, C.~Veaux, and K.~MacDonald, ``Cstr vctk corpus: English
  multi-speaker corpus for cstr voice cloning toolkit (version 0.92),
  [sound],'' https://doi.org/10.7488/ds/2645, 2019.

\bibitem{ljspeech17}
K.~Ito and L.~Johnson, ``The lj speech dataset,''
  https://keithito.com/LJ-Speech-Dataset/, 2017.

\bibitem{BabacanDDHD13}
O.~Babacan, T.~Drugman, and et~al., ``A comparative study of pitch extraction
  algorithms on a large variety of singing sounds,'' in \emph{{IEEE}
  International Conference on Acoustics, Speech and Signal Processing}, 2013,
  pp. 7815--7819.

\bibitem{fastpitch2021}
A.~Łańcucki, ``Fastpitch: Parallel text-to-speech with pitch prediction,'' in
  \emph{2021 IEEE International Conference on Acoustics, Speech and Signal
  Processing}, 2021, pp. 6588--6592.

\end{thebibliography}
\end{document}